# Statistical Failure Mechanism Analysis of Earthquakes Revealing Time Relationships

Parsa Rastin, Michael LuValle

Rutgers University

**If we assume that earthquakes are chaotic, and influenced locally then chaos theory suggests that there should be a temporal association between earthquakes in a local region that should be revealed with statistical examination. To date no strong relationship has been shown (refs not prediction). However, earthquakes are basically failures of structured material systems, and when multiple failure mechanisms are present, prediction of failure is strongly inhibited without first separating the mechanisms. Here we show that by separating earthquakes statistically, based on their central tensor moment structure, along lines first suggested by a separation into mechanisms according to depth of the earthquake, a strong indication of temporal association appears. We show this in earthquakes above 200 Km along the pacific ring of fire, with a positive association in time between earthquakes of the same statistical type and a negative association in time between earthquakes of different types. Whether this can reveal either useful mechanistic information to seismologists, or can result in useful forecasts remains to be seen.**

A number of papers have suggested that earthquakes should be considered chaotic [1-6], and chaos theory suggests that chaotic systems should admit short range statistical prediction [7,8]. This contradicts the general results found in the literature of failed attempts at earthquake prediction [9]. However, from the standpoint of statistical reliability prediction this would make perfect sense if in attempting to predict earthquakes, we are trying to predict failures caused by multiple failure mechanisms without separating the mechanisms.

In 1990 Green and Burnley [10] noticed that very deep earth quakes must be failing from a very different mechanism. Below 100 to 200 kilometers the temperature and pressure are such that brittle fracture can not occur in the materials. Instead a mechanism of phase change in olivine was identified that resulted in an anti-cracking mechanism. Following this lead one of us took data from a central moment tensor data base [11,12] and try to identify the variables in the data base which would best separate deep earthquakes (defined for our purposes as below 200Km) and shallow earthquakes (above 200 Km). This division was chosen because it looked like it would give a reasonable number of earthquakes in each class using a random forest approach [13], 4 variables from the data base were identified. The 4 variables were the azimuth of each of the eigenvectors of the central moment tensor, and the plunge of the third eigenvector. To see if it was possible to simplify the separation, the means in deep and shallow earth quakes along these variables were found, and a vector was constructed using the difference. The earthquakes were projected on the vector. Fitting a simple empirical density plot to deep and shallow earthquakes separately resulted in a clear separation of the earthquakes into 4 distinct groups. The number in each group is shown in table 1 below, and a graph of a density fit (in r using default cross validated window calculations) is shown in figure 1 below on the left. The plot on the right shows a 3 dimensional scatter plot of the earthquakes with the base being latitude and longitude, while the z axis increases with depth in kilometers.

Table1

| Actual depth | Predicted shallow | Predicted deep |
|---|---|---|
| Shallow | 19683 | 16184 |
| Deep | 790 | 2019 |

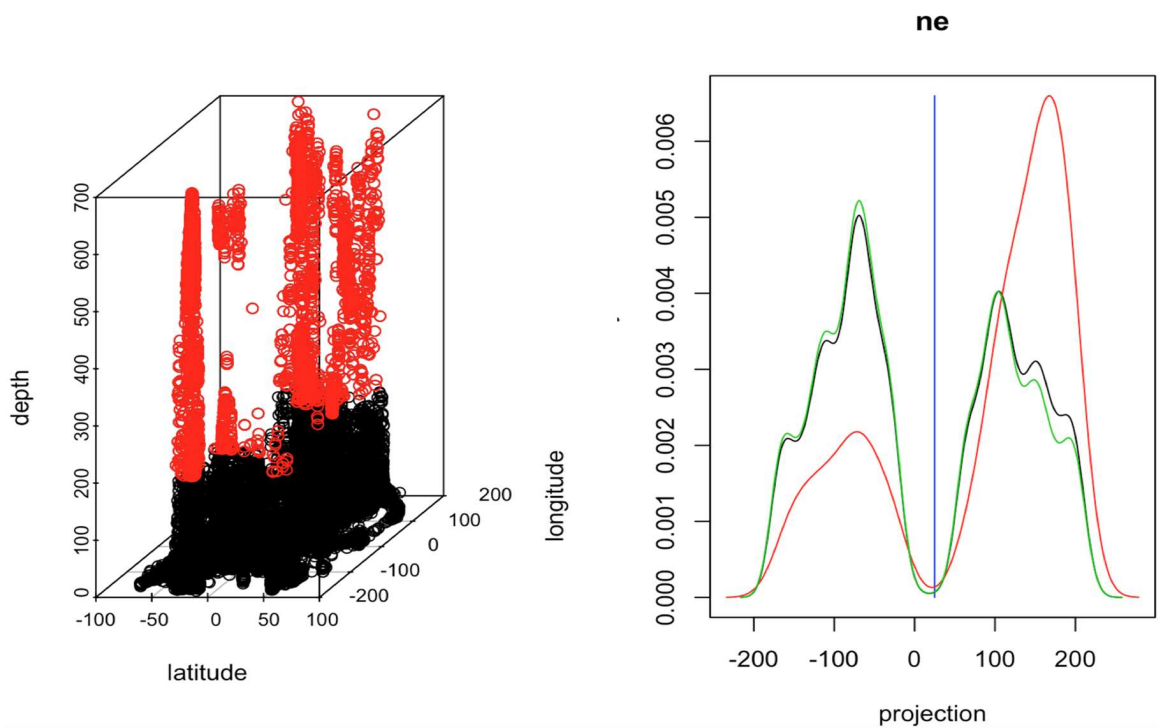

The Green shows the shallow density, the red the deep, each density is calculated with respect to normalizing for the total.

Then using the classification based on splitting each population (shallow and deep) at the blue line, we spit the ring of fire in the pacific into 15 degree by 15 degree "squares" (longitude and latitude), and where both classes of shallow earthquakes existed, we divided time up into approximate 2 week intervals (26 per year) and checked for temporal association with earthquakes (of the same type of earthquake) in the next 2 week period, and did it again skipping a two week period. Everyplace more than 5 of both kinds of earthquakes appeared was checked, and a simple contingency table was built to check temporal independence (see statistical appendix) of the earthquakes within, and across type. We then applied false discovery rate [14] to the statistical tests in each region, and calculated the log odds ratio for each test that was identified as interesting by false discovery rate. The results are shown in figures 2 and 3 below. For control we also checked the test for independence of the shallow earthquakes when there was no attempt in separating them by type. The q-value for the false discovery rate was set at 0.01. An earthquake was counted only when it was larger than 3.05 in magnitude.

Figure 3, Next 2-week period

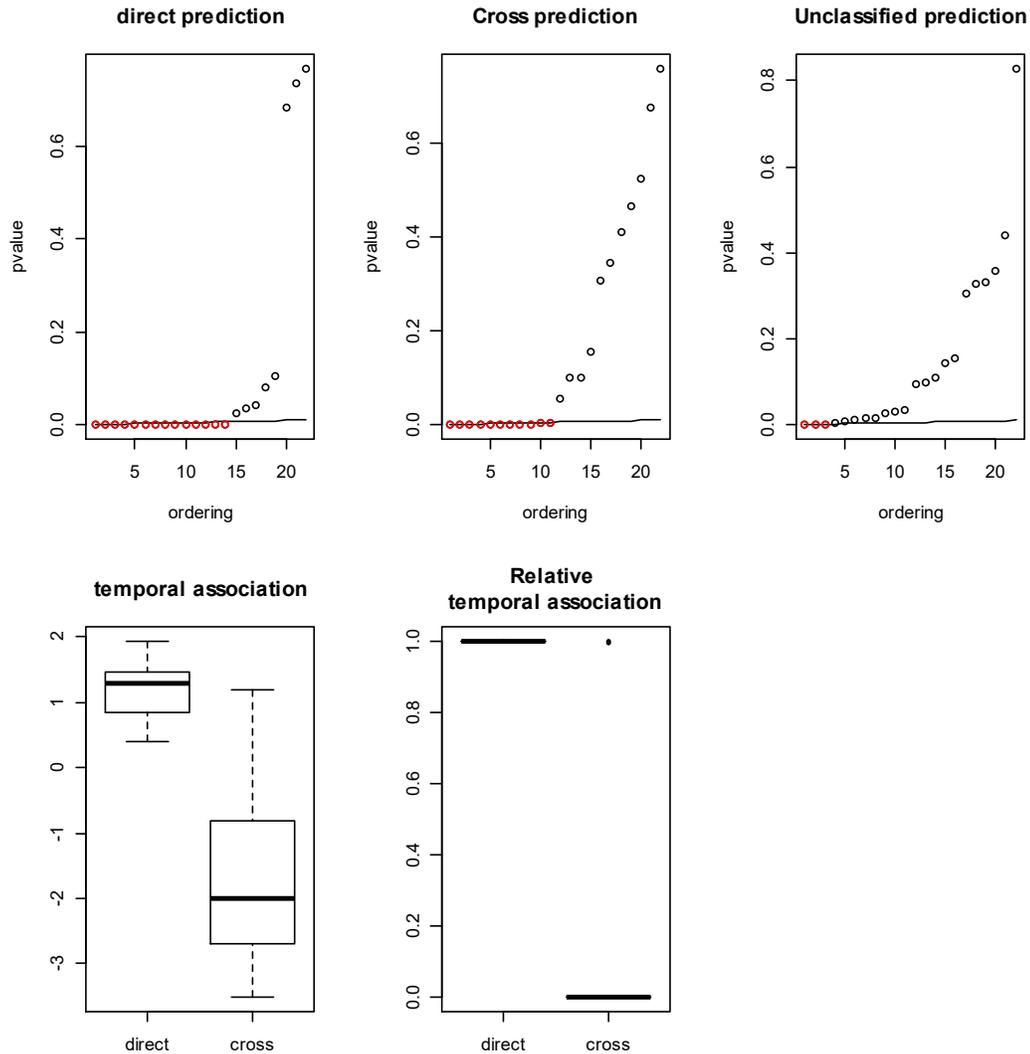

Figure 3 Caption: The plot on the upper left shows the ordered P values for direct comparison within failure modes. The line corresponds to an FDR of 0.01, and the red points are those p-values designated as interesting by the FDR procedure. The plot in the upper Center is the corresponding plot showing ordered p-values across failure modes. The plot in the upper right is the corresponding plot showing when the same procedure is applied to all earthquakes without separating failure modes. The plot on the lower left is a box plot showing the logs odd ratio for all of the p-values chosen as interesting by the FDR procedure in the direct comparison, vs the log odds ratio for each p-value chosen as interesting by the FDR Procedure in the cross failure mode comparison. Roughly, log odds are positive or negative depending on whether there is a positive or negative correlation across time. As discussed in the statistical appendix, the usual interpretations are a bit in question for the contingency tables these are formed from. So the p-values are all calculated using a permutation test for the 3 plots, and the 5[th] plot shows the relative proportion of permuted log-odds ratios lying below each observed log odds ratio

for each case. This relative comparison allows us to see that the temporal association being shown is in fact significant.

Figure 4 Skip forward one 2-week period

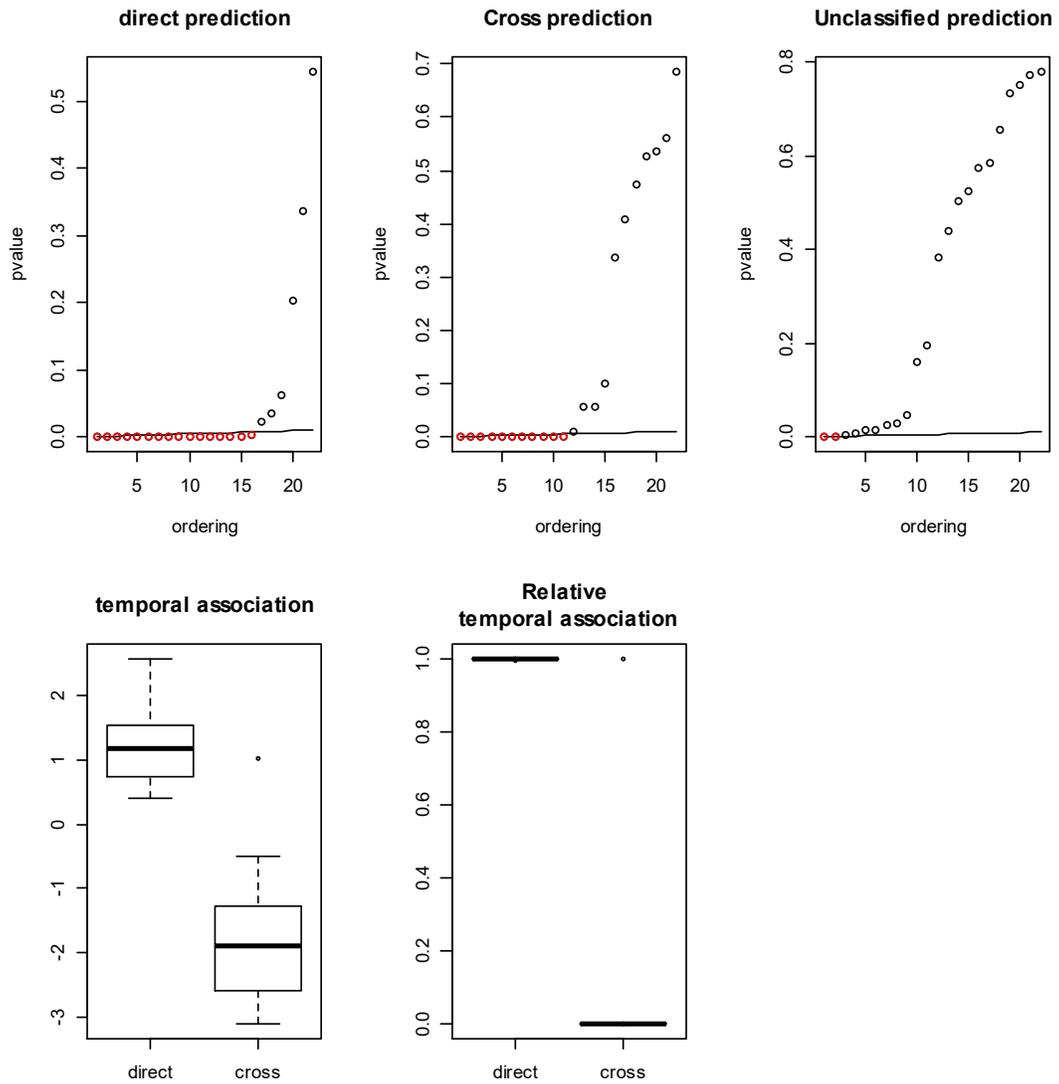

Figure 4 Caption: Same explanation as figure C only the comparison is now across between earthquakes in 2-week periods, separated by one 2-week period.

What we see is separating the shallow earthquake classes into these two failure modes, results in a partition suggesting structure and interaction between faults in each region failing according to the two modes. In particular earthquakes of the same type are likely to follow one another, where as it is less likely that earthquakes of the different type will follow soon after (negative log odds ratio so negative temporal association). If these types can be connected to locations on faults, or perhaps to fault lines, then perhaps a form of forecast based on this interaction can be developed. The interaction between different types suggests the potential for a mechanistic explanation which also may be useful. While what we are observing may simply be aftershocks, the temporal association between earthquakes of the same type, and anti-association between earthquakes of different type is much clearer.

1. Kagan, Y.Y., 1994, *"Observational Evidence of Earthquakes as a Nonlinear Dynamic Process"* Physica D, 160-192
2. Huang, J. and Turcotte, D.Geophys. Res. Lett. 17, 223 (1990)
3. Lacorata, G. and Paladin, G., J. Phys. A 26, 3463 (1993)
4. Field, S., Venturi, N. and Nori, F. (1995), Phys. Rev. Lett. 74, 74
5. de Sousa Vierra, M., (1998), *"Chaos and Synchronized Chaos in an Earthquake Model"*, arXiv CondMat: 9811305v1, Cond-Mat.Stat-Mech
6. Burridge, R., and Knopoff, L., (1967), Bull. Seismol. Soc. Am. 57,341
7. Sauer, T., Yoreck, J. and Casdagli, M. "Embedology", *Journal of Statistical Physics*, **65,** 579-616 (1991)
8. Garland, J. and Bradley E., "*Prediction in projection*", Chaos **25**, 123108 (2015); DOI: 10.1063/1.4936242
9. Geller R.J., Jackson,D.D,, Kagan Y.Y., and Mulargia.F, *"Earthquakes Cannot be Predicted"*, Science V27, 5306 p1616
10. Green, H.W. and Burnley P., (1989), *"A New Self-Organizing Method for Deep Focus Earthquakes"*, Nature, 341, pp 733-737
11. Dziewonski, A. M., T.-A. Chou and J. H. Woodhouse, *"Determination of Earthquake Source Parameters from Waveform Data for Studies of Global and Regional Seismicity"*, J. Geophys. Res., 86, 2825-2852, 1981. DOI: 10.1029/JB086iB04p02825
12. Ekström, G., M. Nettles, and A. M. Dziewonski, "*The global CMT project 2004-2010*": Centroid-moment tensors for 13,017 earthquakes, Phys. Earth Planet. Inter., 200-201, 1-9, 2012. DOI: 10.1016/j.pepi.2012.04.002
13. Fawagreh, F., Gaber, M. M. and Elyan E. (2014) *"Random Forests: From Early Developments to Recent Advancements"*, Systems Science & Control Engineering, 2:1, 602-609, DOI: 10.1080/21642583.2014.956265
14. Benjamini, Y. and Hochberg, Y. (1995), *"Controlling the False Discovery Rate, a Practical and Powerful Approach to Multiple Testing"*, Journal of the Royal Statistical Society*,* Series B, 57, 289-300

Statistical appendix

The approach to checking temporal dependence was to divide time into distinct periods (respectively 26 per year "2-week periods" and 6 per year "2-month periods", in each 15-degree by 15-degree region around the circle of fire plus a little see figure A1 below. For each region examined statistically, the time from 1977 to 2010 was divided into equal time periods, 2 weeks or 2 months. For each type of shallow earthquake, a vector is created. The vector holds a 1 for the 2week period if any earthquakes of that type occur during the time period, and a 0 otherwise. For checking the temporal association, a 1 time period offsite is built into the vector (e.g. The first component is removed from 1 copy of the vector, and the last from a second). Call the vectors for failure mode 1 and 2 respectively $v_1$ and $v_2$ .

$$A = \begin{pmatrix} v_{1,1} \\ ... \\ v_{1,n-1} \\ v_{2,1} \\ ... \\ v_{2,n-1} \end{pmatrix} \quad B = \begin{pmatrix} v_{1,2} \\ ... \\ v_{1,n} \\ v_{22} \\ ... \\ v_{2,n} \end{pmatrix} \quad C = \begin{pmatrix} v_{2,1} \\ ... \\ v_{2,n-1} \\ v_{1,1} \\ ... \\ v_{1,n-1} \end{pmatrix}$$

To create the contingency table for direct temporal association, the two by two table becomes:

Table A1

| Sum(A*B) | Sum(A*(1-B)) |
|---|---|
| Sum((1-A)*B) | Sum((1-A)*(1-B)) |

For cross failure mode temporal association, it takes the form:

Table A2

| Sum(C*B) | Sum(C*(1-B)) |
|---|---|
| Sum((1-C)*B) | Sum((1-C)*(1-B)) |

The entries in the 2X2 tables are not statistically independent, so its not clear that the ordinary distributional properties apply to the natural chi-square statistic and log odds ratio one would create from the table. So instead the p-value of the chi-square statistic was evaluated by permuting each vector randomly and independently 10000 times and reevaluating the chi-square statistic. The P-values used in the FDR calculation are the proportion of these random permutations where the statistic exceeded that arising from the data. The distribution of the log-odds ratios is similarly suspect, so the portion of random log-odds ratios for each vector combination lying below that from the observed was recorded as well resulting in the 5$^{th}$ plot in figures 2 and 3 in the main text.

Figure A1 Tiling pattern

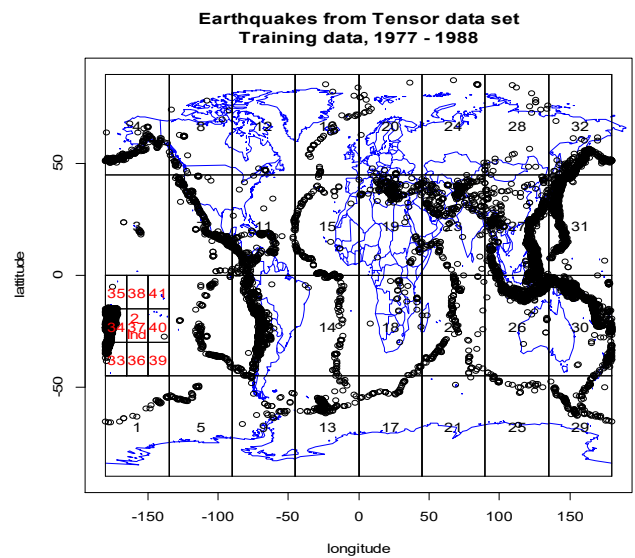

Caption A1: The regions included were 2,10 ,11,7,4,32,31,27,26,30, and 19. Each region was subdivided into 9 sub regions, and those sub regions where more than 5 shallow earthquakes of each of the two types were examined statistically.

As mentioned, the calculations were also done for 2 months ahead, and 2 months ahead, skipping 2 months with the plots shown below corresponding to figures 2 and 3 respectively in the main text.

Figure A 2 Next 2-month period

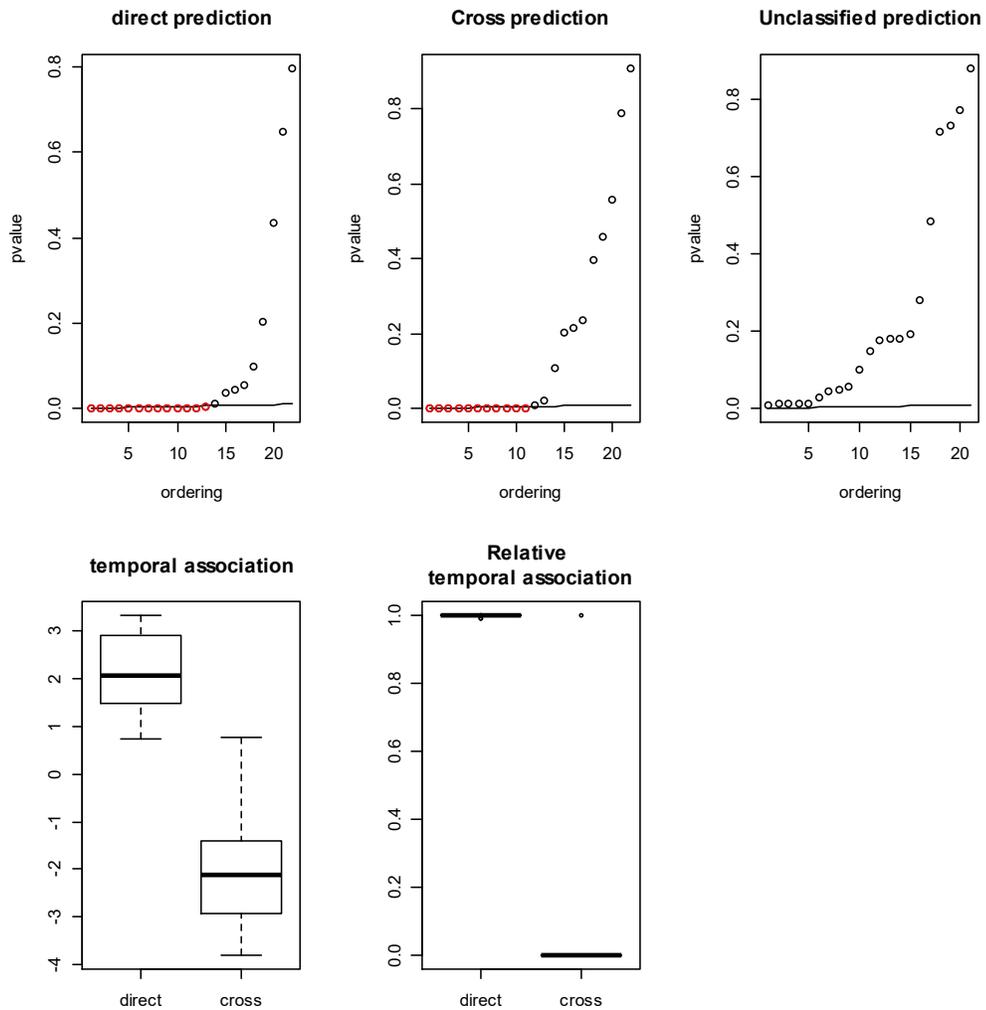

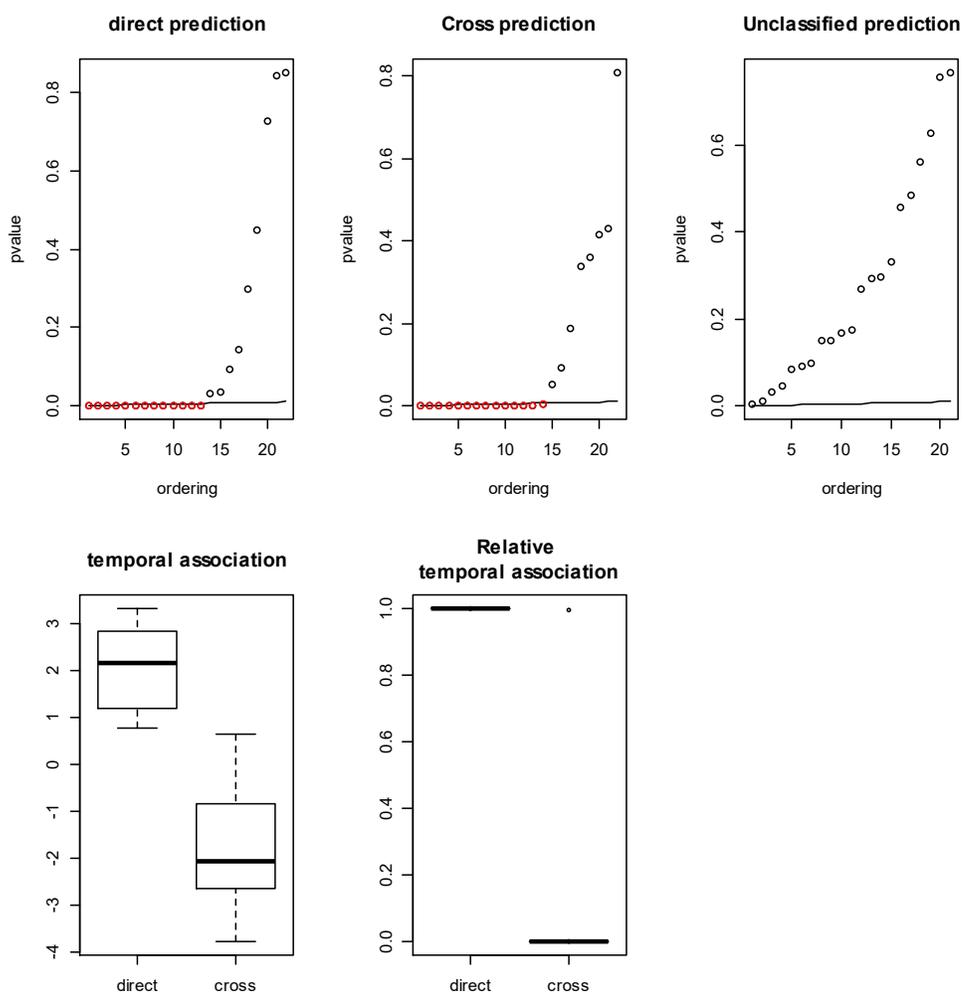

Figure A3: 2-month period, skipping 1 2-month period